\documentstyle[prd,aps,floats]{revtex}
\begin{document}
\draft
\preprint{SUSSEX-AST 98/6-1, astro-ph/9806127}
 
%
%
\input epsf
\renewcommand{\topfraction}{0.8}
\twocolumn[\hsize\textwidth\columnwidth\hsize\csname
@twocolumnfalse\endcsname
 
\title{Inflation during oscillations of the inflaton}
\author{Andrew R.~Liddle and Anupam Mazumdar}
\address{Astronomy Centre, University of Sussex, Falmer, Brighton BN1
9QJ,~~~U.~K.}
\date{\today}
\maketitle
\begin{abstract}
Damour and Mukhanov have recently devised circumstances in which
inflation may continue during the oscillatory phase which ensues once
the inflaton field reaches the minimum of its potential. We confirm the 
existence of this phenomenon by numerical integration. In such circumstances 
the quantification of the amount of inflation requires particular care. We
use a definition based on the decrease of the comoving Hubble length,
and show that Damour and Mukhanov overestimated the amount of
inflation occurring. We use the numerical calculations to check the
validity of analytic approximations.
\end{abstract}
 
\pacs{PACS numbers: 98.80.Cq \hspace*{2.1cm} Sussex preprint SUSSEX-AST
98/6-1, astro-ph/9806127}
 
\vskip2pc]
 
 
\section{Introduction}

Ordinarily, cosmological inflation --- a period of accelerated
expansion --- is considered to be driven by a scalar field $\phi$
rolling slowly, and monotonically, down a shallow potential
$V(\phi)$. However, recently Damour and Mukhanov \cite{DM} have
pointed out that for non-convex potentials, where $d^2V/d\phi^2$ is
negative in regions not too far from the minimum, there exist
circumstances where inflation may continue during the
oscillations. The basic idea is to arrange that the scalar field
spends most of its time of the shallow ``wings'' of the potential,
where it is quite flat, so that for each oscillation there is a period
of inflation which overpowers the inevitable non-inflationary region
near the core of the potential.  One thus obtains on average some
inflation over a complete cycle of oscillation. Damour and Mukhanov
did not give such behaviour a name; we shall call it {\em oscillating
inflation}.

The oscillations take place on a much shorter timescale than the
Hubble expansion. However, over many oscillations the effect of the
expansion is felt, and drains energy away from the oscillations.  This
process continues until the scalar field, known as the inflaton, is
trapped completely within the convex core of the potential. There the
condition for oscillating inflation fails and inflation ceases in the
usual manner.

Because each oscillation is a combination of inflating and
non-inflating portions, considerable care is required in defining what
is meant by the amount of inflation obtained, and in fact Damour and
Mukhanov overestimated the amount of inflation taking place. We
discuss in detail the definition of the amount of inflation, and
confirm our analytic estimates numerically. 


\section{Dynamics}

In this section we briefly summarize the results obtained by Damour
and Mukhanov \cite{DM}.  The evolution equations of the scalar field
for the flat Friedmann cosmology read 
\begin{eqnarray}
\label{DYN1}
\ddot\phi + 3H\dot\phi & = & - \frac{dV}{d\phi} \,;\\
H^{2} &=& \frac{8\pi}{3m_{{\rm Pl}}^2} \rho \,,
\end{eqnarray}
where $H$ is the Hubble parameter and $\rho \equiv \frac{1}{2}\dot
\phi^2 + V(\phi)$ is the energy density of scalar field. Its pressure
$p$ is given by $p \equiv \frac{1}{2}\dot \phi^2- V(\phi)$. {}From
these two further equations can be obtained, which are useful though
of course not independent of the first two, namely
\begin{eqnarray}
\dot \rho &=& - 3H(\rho + p) = -3H\dot \phi^2 \,;\\
\label{Acc}
\frac{\ddot a}{a} &=& -\frac{4\pi}{3 m_{{\rm Pl}}^2}(\rho + 3p) \,.
\end{eqnarray}

The effective adiabatic index of the scalar matter can be defined as
\begin{equation}
\gamma \equiv \frac{\rho+p}{\rho} = \frac{\dot \phi^2}{\rho} 
	\approx \frac{2}{3}\epsilon_{{\scriptscriptstyle H}} \,.
\end{equation}
where $\epsilon_{{\scriptscriptstyle H}}$ is the slow-roll parameter,
defined as in Ref.~\cite{LPB}, which is required to be less than one
for inflation to proceed.  Inflation occurs whenever $\gamma < 2/3$,
corresponding to $\rho + 3p < 0$.

We assume that initially the scalar field is well displaced from the
minimum, and drives a period of slow-roll inflation. We will also
follow Damour and Mukhanov in assuming the potential is an even
function of $\phi$. As the field approaches the minimum, eventually
the slow-roll conditions cease to apply, at some $\phi$ which depends
on the nature of the potential. Then rapid oscillations, with
frequency $\omega$, dominate the evolution equations. For $\omega \gg
H$, one is dealing with two timescales; individual oscillations can
be studied ignoring the Hubble expansion, and the effect of the
expansion imposed on the behaviour averaged over oscillations.
Ignoring the Hubble expansion for the time being, the amplitude of the
oscillations remains approximately constant with $\rho \approx
V(\phi_{{\rm m}})$, $\phi_{{\rm m}}$ being the highest point of
oscillation. Eq.~(\ref{DYN1}) then reduces to  
\begin{equation}
\label{MDYN}
\frac{d}{dt} \left( \frac{1}{2} \dot \phi^{2} + V(\phi) \right) = 0 \,; \\
\end{equation}
and the period $\tau$ of such oscillations is obtained by integrating
Eq.~(\ref{MDYN}).
\begin{equation}
\tau \equiv \frac{2\pi}{\omega} = 4 \int_{0}^{\phi_{{\rm m}}} \frac{d\phi} 
	{\sqrt{2\left[ V_{{\rm m}} - V(\phi) \right]}} \,,
\end{equation}
where $V_{{\rm m}} = V(\phi_m)$. We obtain the adiabatic index by
averaging over an oscillation
\begin{equation}
\gamma\equiv \left\langle \frac{ \rho + p }{\rho} \right\rangle \equiv
	\left\langle \frac{\dot\phi^2}{\rho} \right\rangle \,.
\end{equation}
The condition $\gamma < 2/3$ can be expressed as 
\begin{eqnarray}
\label{EQUAL}
\gamma &=&\frac{\langle \dot \phi^{2}\rangle}{V_{{\rm m}}} 
=2\left(1-\frac{\langle V(\phi)\rangle}{V_{{\rm m}}} \right) \,
\nonumber \\ &=&2\frac{\int_{0}^{\phi_{{\rm m}}}(1- V(\phi)/V_{{\rm
m}})^{1/2}d\phi} {\int_{0}^{\phi_{{\rm m}}}(1- V(\phi)/V_{{\rm
m}})^{-1/2}d\phi} < \frac{2}{3} \,.
\end{eqnarray}
Using first two equalities one can reduce this inequality to a simpler form
\begin{equation}
\label{cond}
\langle V - \phi V_{,\phi} \rangle > 0 \,,  
\end{equation}
where the comma indicates a derivative.

This has a nice geometric interpretation \cite{DM}; $V-\phi V_{,\phi}$
is the intercept of the tangent to the potential at the point $\phi$,
and to obtain inflation on average, then over an oscillation the
intercept has to be positive.  Note that the above condition fails to
be satisfied for power-law potentials ($V(\phi) \propto \phi^{q}$) for
$q \geq 1$. Such potentials must be a good approximation sufficiently
close to the minimum of any potential.

\section{Quantifying inflationary expansion}

In slow-roll inflation, the standard quantification of inflation is
the number of $e$-foldings $N$, defined by
\begin{equation}
N = \ln \frac{a_{{\rm f}}}{a_{{\rm i}}} \,,
\end{equation}
where `i' and `f' denote initial and final values respectively.
However, if slow-roll is not working well this requires
modification. This is particularly true for oscillating inflation,
where the universe continues to expand during those non-inflationary
parts of the expansion. They should certainly not count towards the
inflationary total, and in fact should count {\em against} it.

In such circumstances, the correct definition is to examine the change
in the comoving Hubble length \cite{LPB}, $H^{-1}/a$, by defining
\begin{equation} 
\label{num}
\tilde{N} = \ln \frac {a_{{\rm f}} H_{{\rm f}}}{a_{{\rm i}} 
	H_{{\rm i}}} \,. 
\end{equation}
There are many reasons why this is the correct general definition.
Firstly, the inflationary condition $\ddot{a} > 0$ is precisely the
condition that $aH$ is increasing. Secondly, it is the combination
$aH$ which determines whether the flatness and horizon problems are
being solved.  And, most importantly of all, the condition for horizon
crossing for a perturbation of comoving wavenumber $k$ is $k=aH$, so
it is that combination which decides whether scales are inside or
outside the horizon.

In the slow-roll limit $N$ and $\tilde{N}$ are identical, but here
they are far from it, and $\tilde{N} < N$ by definition as $H$ always
decreases. 

\section{The Damour--Mukhanov model}

\subsection{Analytic approximation}

So far the argument holds good for arbitrary potentials. {}From now on
we shall restrict ourselves to potentials which give rise to inflation
during oscillations. We begin by considering the one suggested by
Damour and Mukhanov \cite{DM}, which is the potential
\begin{equation}
\label{pot}
V(\phi)=\frac{A}{q}\left[\left( \frac{\phi^2}{\phi_{{\rm c}}^2} + 1
	\right)^{q/2} -1 \right] \,,
\end{equation}
where $q$ is a real dimensionless parameter greater than zero. $A$ and
$\phi_{{\rm c}}$ are dimensionful; $A$ has the dimension of [mass]$^4$
and $\phi_{{\rm c}}$ is the scale which determines the core of the
potential and has the same dimension as $\phi$, namely [mass]. The potential 
is shown in Fig.~1. It can be reduced to different forms in different 
regimes. For
$q \rightarrow 0$, $V(\phi) = \frac{1}{2} A \, \ln( 1 + \phi^2/\phi_{{\rm
c}}^2)$, while when $\phi_{{\rm c}} \ll \phi$ the potential reduces to
$V(\phi)\approx Aq^{-1} \left(\phi/\phi_{{\rm c}} \right)^q $. If both
limits are taken one gets $V(\phi) \approx A\ln(\phi/\phi_{{\rm c}})$.

\begin{figure}[t]
\centering
\leavevmode\epsfysize=6cm \epsfbox{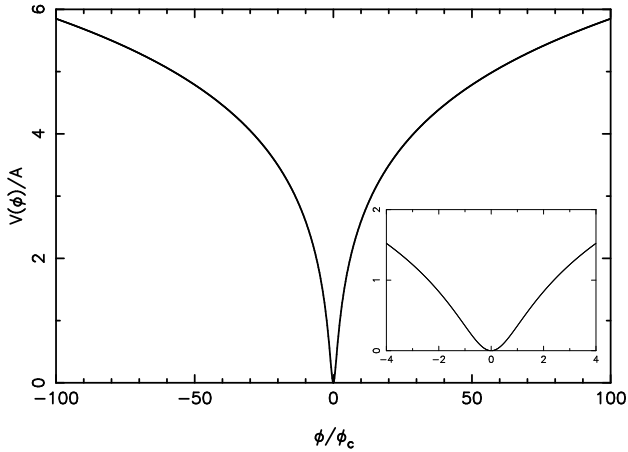}\\
\caption[fig1]{The potential for $q=0.1$, showing the concave shape for 
$|\phi| \gg \phi_{{\rm c}}$. The insert shows the convex region around the 
minimum.}
\end{figure}

Since oscillating inflation will only occur if the oscillations extend well
outside the core region, for most of a cycle the field must obey $\phi \gg
\phi_{\rm c}$ so that the potential reduces to the power-law form.  We can
approximate the mean behaviour by ignoring the core region, an approximation
which will hold well until the oscillation amplitude falls close to the core
radius and oscillating inflation ceases.  For power-law potentials the
adiabatic index $\gamma$ has been computed in Ref.~\cite{Turner} as $\gamma =
2q/(q+2)$ and the various physical quantities evolve as \cite{DM}
\begin{eqnarray}
\label{evol1}
a &\propto& t^{2/{3\gamma}} = t^{(q+2)/3q} \, \\
\label{evol2}
\rho = V(\phi_{{\rm m}}) &\propto& t^{-2} \propto a^{-6q/(q+2)} \, \\
\label{evol3}
\phi_{{\rm m}} &\propto & t^{{-2}/{q}} \propto a^{-6/(q+2)}. 
\end{eqnarray}
where $\phi_{{\rm m}}$ is the amplitude of the oscillations,
$\phi_{{\rm c}} < \phi_{{\rm m}} < \phi_{{\rm s}}$, $\phi_{{\rm s}}$
being the value of $\phi$ where the slow-roll approximation breaks
down.

\begin{figure}[t]
\centering
\leavevmode\epsfysize=6cm \epsfbox{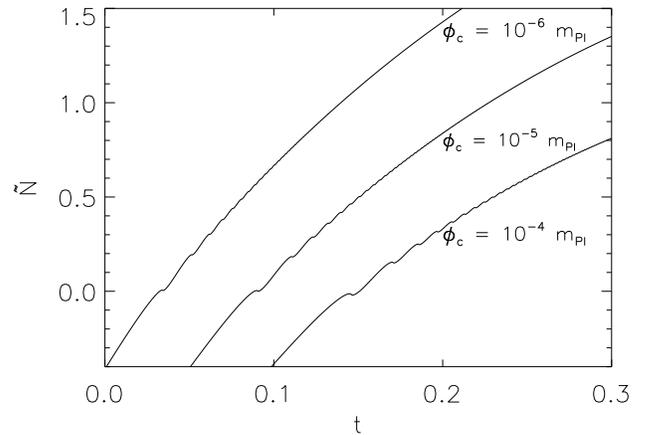}\\
\leavevmode\epsfysize=6cm \epsfbox{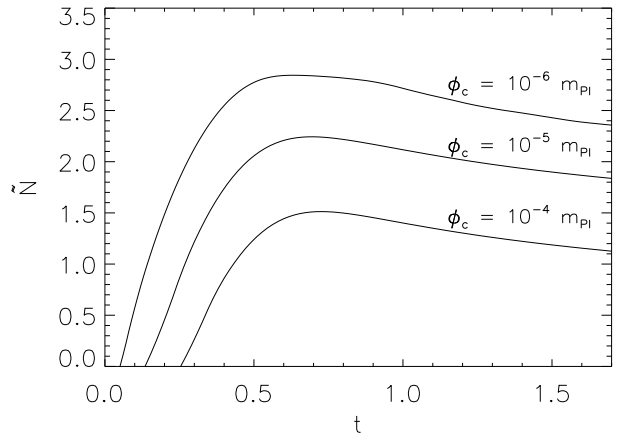}\\
\caption[fig2]{A numerical simulation of the evolution of $\tilde N$ for
$q=0.1$. The upper panel shows the early evolution for two choices of 
$\phi_{{\rm c}}$, showing the last stages of slow-roll and the beginning of 
the oscillations. The lower panel shows the 
complete oscillating inflation era for two choices of $\phi_{{\rm c}}$; the 
oscillations are too small to be seen on this scale. In 
each case, $\tilde{N}$ is normalized to zero at the end of slow-roll 
inflation, and an arbitrary horizontal shift has been used to separate the 
lines. The heights of the maxima in the lower panel are accurately given by 
Eq.~(\ref{num3}).}
\end{figure}

{}From Eqs.~(\ref{evol1}-\ref{evol3}) one can compute Eq.~(\ref{num})
\begin{equation}
\label{num1}
\tilde{N} \simeq \frac{1-q}{3} \ln 
	\frac{\phi_{{\rm s}}}{\phi_{{\rm c}}} \,,
\end{equation}
where $\phi_{{\rm s}}$ is the value of the field at the end of slow
roll, which is usually less than $m_{{\rm Pl}}$. This is to be
contrasted with the amount of expansion 
\begin{equation}
\label{num2}
N \simeq \frac{2+q}{6} \ln \frac{\phi_{{\rm s}}}{\phi_{{\rm c}}} \,.
\end{equation} 
Damour and Mukhanov \cite{DM} quoted this expression, with the
additional approximation $\phi_{{\rm s}} \simeq m_{{\rm Pl}}$.  This
latter result is quite misleading, for example it suggests that there
is a possibility of enhancement of accelerated expansion due to
increase in the value of $q$. Their result is self-contradictory, as
we have already seen that there can be no inflation during
oscillations for the power-law potentials with $q \geq 1$. Our result
is supported by the numerical calculation, described in the next
subsection.

We also note that the assumption $\phi_{{\rm s}} \simeq m_{{\rm Pl}}$
is not correct in general. The contrived shape of the potential allows
field to slow roll for a longer period. {}From Eq.~(\ref{num1}), we see
that as we approach $q \rightarrow 0$ the prefactor increases and one
expects to get higher values of $\tilde{N}$ for the same values of
$\phi_{{\rm c}}$. In fact, on the contrary one gets smaller
$\tilde{N}$, because slow roll does not end at $\phi_{{\rm s}} \approx
m_{{\rm Pl}}$ but rather at $\phi_{{\rm s}} \approx q m_{{\rm
Pl}}/\sqrt{16\pi}$ for $\phi \gg \phi_{{\rm c}}$, giving
\begin{equation}
\label{num3}
\tilde{N} \simeq \frac{1-q}{3} \left[ \ln 
	\frac{q \, m_{{\rm Pl}}}{\phi_{{\rm c}}} - 2 \right] \,.
\end{equation}
As $q$ approaches zero, $\phi_{{\rm s}}$ also goes to zero, implying
that the number of $e$-foldings in Eq.~(\ref{num1}) also decreases
logarithmically. This effect starts dominating as $q \rightarrow 0$
lowering the maximum yield in $\tilde N$, determined by Eq.~(\ref{num3}).
Hence to obtain the maximum number of $e$-foldings
one needs to fix the value of $q$ and then reduce $\phi_{{\rm c}}$.

For sensible values of the parameters, only a few $e$-foldings
of inflation are available. Even if $\phi_{{\rm c}}$ is lowered to the
electro-weak scale, $\phi_c= 10^{-17} m_{\rm Pl}$, the amount is still
limited, for example $q=0.6$ gives $\tilde N \approx 4.9$ and $q=0.1$
gives $\tilde N \approx 10.5$. 

For a logarithmic potential the adiabatic index $\gamma = 1/\ln
(\phi_{{\rm m}} /\phi_{{\rm c}})$ \cite{DM}, and we are unable to give
a simple form of $\tilde{N}$ for such a potential.

\begin{figure}[t]
\centering
\leavevmode\epsfysize=6cm \epsfbox{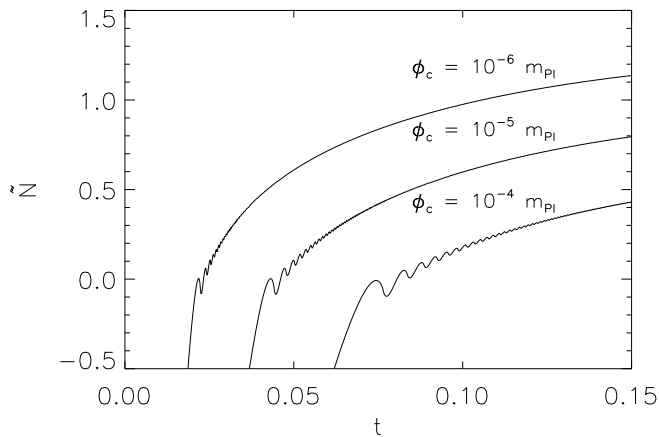}\\
\caption[fig3]{As the upper panel of Fig.~2, but for $q=0.6$. }
\end{figure}

\subsection{Numerical analysis}

In this section we describe our numerical results. We consider the 
potential of Eq.~(\ref{pot}) and evolve the field equations numerically. In 
Figs.~2 and 3
we depict the evolution of $\tilde N$ for two different values
of $q$. The first hump describes the end of slow-roll approximation,
which is where oscillating inflation is considered to start. For
normal potentials there is a sharp decrease in the value of $\tilde N$
after slow-roll inflation ends, but for the potential under
consideration we see further $e$-foldings; each oscillation has
inflationary and non-inflationary parts but the former dominate
leading to a continued upward trend. Eventually, as seen in the lower
panel of Fig.~2, the field finds its way to the true core and
oscillating inflation ends. Its total amount is measured by the maximum
height of $\tilde{N}$ above its value at the end of slow-roll
inflation.

Fig.~4 compares our analytic estimate Eq.~(\ref{num3}) of the amount
of inflation with our numerical results, as a function of $q$.  The
agreement is extremely good.  There is no inflation for $q > 1$, and
as $q$ is decreased the amount of inflation increases almost linearly
before reaching its maxima and then starts decreasing as $q$
approaches zero.  In this last regime the analytic approximation fails
to work well, with the potential approaching the logarithmic limit as
explained in the previous section.  For small values of $q$ we are
unable to compare the numerical results as we have not been able to
derive any analytical expression.

\begin{figure}[t]
\centering
\leavevmode\epsfysize=6cm \epsfbox{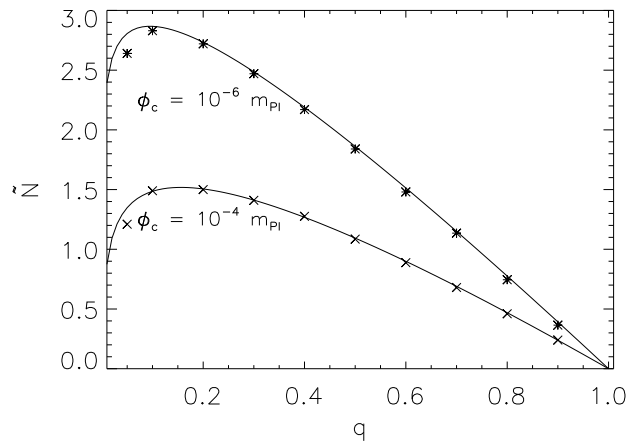}\\
\caption[fig4]{A comparison of numerical and analytical solutions
of the number of e-foldings for $\phi_c =10^{-4} m_{\rm Pl}$ and
$10^{-6}m_{\rm Pl}$. The symbols correspond to the numerical results,
and the smooth curves to the analytical estimate of Eq.~(\ref{num3}). }
\end{figure}

\section{Density perturbations}

As well as solving problems of initial conditions, inflation plays a
crucial role in generating density perturbations which can later seed
structure formation. Because the oscillating inflation is brief, the observed 
perturbations are normally expected to originate in the slow-roll regime 
which precedes the oscillations, though in principle a subsequent additional 
period of inflation could push them to observable scales \cite{DM}. In this 
short section, we check that required amplitude of density perturbations 
generated in the slow-roll epoch does not give an uncomfortable constraint on 
the potential under consideration.

Defining the primordial density perturbation
spectrum $\delta_{{\rm H}}$ as in Ref.~\cite{LL}, one has
\begin{equation}
\delta_{{\rm H}}^2 \approx \frac{32}{75} \, \frac{V}{m_{{\rm Pl}}^4}
	\, \frac{1}{\epsilon} \,,
\end{equation}
where the slow-roll parameter $\epsilon$ is given by
\begin{equation}
\epsilon = \frac{m_{{\rm Pl}}^2}{16 \pi} \, \left( \frac{V_{,\phi}}{V}
	\right)^2 \,.
\end{equation}
The perturbations observed by the COBE satellite require $\delta_{{\rm
H}} \approx 2 \times 10^{-5}$; since the number of $e$-foldings during
the oscillating inflation phase is small, these perturbations must be
generated during the slow-roll phase. This phase ends when $\epsilon
\approx 1$, which as commented above is at
\begin{equation}
\phi_{{\rm s}} \approx \frac{q}{\sqrt{16\pi}} \, m_{{\rm Pl}} \,.
\end{equation}
During the slow-roll phase the power-law approximation to the
potential is an excellent one and yields
\begin{equation}
\delta_{{\rm H}}^2 = \frac{512\pi}{75} \, \frac{A}{q^3 m_{{\rm Pl}}^6}
	\, \frac{\phi^{2+q}}{\phi_{{\rm c}}^q} \,.
\end{equation}

The value of the amplitude $A$ of the potential is to be adjusted to give the
right level of perturbations when our present Hubble scale crossed outside
the horizon during inflation, of order 50 $e$-foldings before inflation
finally ends. The amount of inflation in the slow-roll epoch can be 
adequately computed using the normal slow-roll formula 
\begin{equation}
N \simeq - \frac{8\pi}{m_{{\rm Pl}}^2} \int_\phi^{\phi_{{\rm s}}}
	\frac{V}{V_{,\phi}} \, d\phi \,,
\end{equation}
and approximating $\tilde{N} \simeq N$. We assume the perturbations are 
generated 50 $e$-foldings from the end of inflation, and ignore the 
oscillating inflation contribution as negligible given the uncertainty in 
this number. We then find, for example, that with 
$q=0.1$ and $\phi_{{\rm c}} = 10^{-6}
m_{{\rm Pl}}$, the required amplitude of the potential is $A^{1/4} \simeq 5
\times 10^{-3} m_{{\rm Pl}}$. This is a fairly typical sort of number for
inflationary models.  We conclude that the appropriate level of density 
perturbations can be readily achieved in these models.

\section{Summary}

We have analyzed the possibility of oscillating inflation both
analytically and numerically. A more accurate quantification of the
amount of inflation shows that it was overestimated by Damour and
Mukhanov \cite{DM}, and we have given a more appropriate definition
which is in excellent agreement with our numerical analysis. Only a
few $e$-foldings of oscillating inflation are possible.

\acknowledgments
A.R.L.~is supported by the Royal Society and A.M.~by the INLAKS
foundation and ORS. ARL thanks the Fermilab astrophysics
group, CfPA Berkeley and the University of Illinois at
Urbana--Champaign for hospitality while part of this work was done. We
acknowledge use of the Starlink computer system at the University of
Sussex.

\end{document}